\documentclass{elsart}
\usepackage{amssymb}

\usepackage{graphicx}
\usepackage{amsmath}

\journal{Physics Letters A}

\begin{document}

\begin{frontmatter}
\title{Tomograms in the Quantum-Classical transition}
\author[rus]{V.I. Man'ko\corauthref{cor1}},
\ead{manko@na.infn.it}
\author[na]{G. Marmo}, \ead{marmo@na.infn.it}
\author[na]{A. Simoni\corauthref{cor1}},
\corauth[cor1]{Corresponding authors} \ead{simoni@na.infn.it}
\author[bama]{A. Stern}, \ead{astern@bama.ua.edu}
\author[na]{F. Ventriglia}\ead{ventriglia@na.infn.it}
\address[rus]{ P.N.Lebedev Physical Institute, Leninskii Prospect 53, Moscow 119991, Russia}
\address[na]{Dip. Sc. Fisiche dell'Universit\`{a} Federico II e Sez. INFN di Napoli, \\ Compl. Univ. Monte S.Angelo, I-80126 Naples, Italy}
\address[bama]{Department of Physics, University of Alabama, Tuscaloosa, AL 35487, USA}

\begin{abstract}

The quantum-classical limits for quantum tomograms are studied and
compared with the corresponding classical tomograms, using two
different definitions for the limit. One is the Planck limit where $\hbar \rightarrow 0$ in all $\hbar -$%
dependent physical observables, and the other is the Ehrenfest limit where $%
\hbar \rightarrow 0$ while keeping constant the mean value of the
energy. The Ehrenfest limit of eigenstate tomograms for a particle
in a box and a harmonic oscillator is shown to agree with the
corresponding classical tomograms of phase-space distributions,
after a time averaging. The Planck limit of superposition state
tomograms of the harmonic oscillator demonstrates the decreasing
contribution of interference terms as $\hbar \rightarrow 0$.
\end{abstract}
\end{frontmatter}

\section{Introduction}

The quantum-classical transition has been the subject of numerous studies,
\cite{MarSud....,Dod....,gutz} for the classical limit of bi-Hamiltonian
systems see e.g. \cite{Alberto....}. In the standard formulation of quantum
mechanics the states of a system are associated with wave functions\cite
{Schroe} or density operators,\cite{Landau27,vonNeum27} and the WKB
procedure can be employed to take the classical limit. Further insight in
the connection to classical mechanics can be gained from other formulations
of quantum mechanics,\cite{AmJ.Phys} e.g., the Wigner-Moyal\cite{Wig32,Moy49}
and Feynman path integral formulations.\cite{Feyn48} In the Moyal approach,
the Wigner function satisfies the quantum evolution equation which is
similar to the kinetic equation for classical probability distributions in
phase-space. In the limit $\hbar \rightarrow 0$ the Moyal equation yields
the Liouville equation of classical statistical mechanics. The classical
action plays an important role in the Feynman path integral representation
for the Green functions (propagators) of the Schroedinger equation, and the
only contributions that appear in the complex probability in the $\hbar
\rightarrow 0$ limit are associated with classical paths. Also the classical
action satisfying the classical Hamilton-Jacobi equation appears in the wave
function after using the WKB decomposition.

Despite all the utility of these well known approaches, because objects such
as wave functions, Wigner functions or other representations of density
operators(e.g., Sudarshan-Glauber singular P-quasidistribution\cite
{Sud63,Glau63} and Husimi-Kano Q-quasidistribution\cite{Hus40,Kano56})
differ in an essential way from classical phase-space probability
distributions, it is not easy to make a precise comparison between the two
systems. This can be rectified by going to the tomographic approach where
both the classical and quantum theory offer a discription in terms of
tomographic probabilities.\cite{Olga97,MendesPhisicD,Mancini} The quantum
tomogram realizes a specific version of star-product quantization\cite
{marmoJPA} and allows for a comparison with the classical tomogram, both
being written on the same domain. The advantage of a tomogram is connected
with its property to be a standard positive probability distribution
function describing the quantum state and this property was found useful to
study the entanglement criterion for continuous variables,\cite{sudarPLA} as
well as to formulate a separability condition for spin degrees of freedom.
\cite{sudarJOB}

The aim of this work is to consider the quantum-classical transition within
the framework of the tomographic approach. We shall mainly be concerned with
examples and postpone more general considerations to a future paper. In
particular, we shall consider the tomographic approach for the description
of quantum states for a particle in a box, as well as the quantum tomograms
of harmonic oscillator coherent states and stationary states. We study two
kinds of quantum-classical transitions, the Planck limit and Ehrenfest limit
of the quantum tomograms. In Planck limit we obtain the quantum tomograms of
harmonic oscillator coherent states and stationary states for $\hbar
\rightarrow 0$. In the Erhenfest limit we obtain the limit of the quantum
tomogram of harmonic oscillator stationary states and of the states of a
particle in a box for $\hbar \rightarrow 0$ with fixed energy. We will show
that the Ehrenfest limit provides the time-average of classical motion (both
for the harmonic oscillator and the particle in a box). Also we study the
behaviour of tomograms for superposition states of the harmonic oscillator
in the Planck limit. Namely, we show that the tomogram of the superposition
of two arbitrary eigenstates (the Schroedinger cat\cite{Schroedinger35}),
goes to a mixture of classical tomograms of these states. The interference
contribution to the tomogram goes to zero in the Planck limit. Another type
of Schroedinger cat system (consisting of even and odd coherent states\cite
{Phisica74}) will be considered in both the Planck and Ehrenfest limits.

The article is organized as follows: in section 2 we review known properties
of symplectic tomograms for classical and quantum systems. In section 3 we
introduce the quantum-classical transition. In subsection 3.1 the Planck
limits are computed for few systems. In subsection 3.2 we study the
Ehrenfest classical limit for the stationary states of the harmonic
oscillator and the infinite square well potential. Concluding remarks and
perspectives are given in the final section. Some relevant formulae for
generalized functions and their limits are considered in the appendix.

\section{Symplectic tomograms in classical and quantum mechanics}

Following references\cite{Olga97,MendesPhisicD,Olga2004} we review the
tomographic description of particle states in classical and quantum
mechanics. For simplicity we restrict to one-dimensional particle systems.

\subsection{Classical tomograms}

The standard description of classical states with fluctuations is given by a
non-negative joint probability distribution function $f(p,q)$ on the phase
space of the particle with one degree of freedom. The function is
normalized, i.e.
\begin{equation}
\int f(p,q)dpdq=1.
\end{equation}
The tomogram is the probability distribution function in a rotated and
scaled reference frame on the phase-space. The classical tomogram is
constructed from $f(p,q)$ and is a function of a coordinate $X\in \mathbb{R}$%
, which is related to the position $q$ and momentum $p$ of a canonical
reference frame on phase-space by
\begin{equation}
X=\mu q+\nu p,  \label{metasimp}
\end{equation}
where $\mu $ and $\nu $ are real parameters. It can be expressed in terms of
a scaling parameter $s$ and a rotation parameter $\theta :$%
\begin{equation}
\mu =s\cos \theta \ ,\ \nu =s^{-1}\sin \theta .
\end{equation}
For fixed $\mu $ and $\nu $ one then gets a line on the commutative plane $%
(q,p)$ with an orientation $\theta $ from the position axis. Thus the
physical meaning of the variable $X$ is that it is the ``position'' of the
particle measured in the reference frame of the phase-space whose axes are
rotated by an angle $\theta $ with respect to the old reference frame, after
preliminary canonical scaling of the initial position $q\rightarrow sq$ and
momentum $p\rightarrow s^{-1}p$. The coordinate $X$ of equation (\ref
{metasimp}) together with
\begin{equation}
Y=-s^{2}\nu q+s^{-2}\mu p
\end{equation}
provides a canonical transformation preserving the symplectic form in the
phase-space. For that reason the classical tomogram is called ``symplectic''.

The tomogram of the classical statistical density $f(p,q)$ is defined by the
Radon transform\cite{gelfand}
\begin{equation}
\mathcal{W}(X,\mu ,\nu )=\int f(p,q)\delta (X-\mu q-\nu p)dpdq.
\label{tom classico}
\end{equation}
It can be written in the form
\begin{equation}
\mathcal{W}(X,\mu ,\nu )=\left\langle \delta (X-\mu q-\nu p)\right\rangle
_{f}  \label{tom'}
\end{equation}
where the average is done using the probability distribution $f(p,q)$ in the
phase space. Its\ inverse is
\begin{equation}
f(p,q)=\frac{1}{\left( 2\pi \right) ^{2}}\int \mathcal{W}(X,\mu ,\nu )\exp %
\left[ i(X-\mu q-\nu p)\right] dXd\mu d\nu .  \label{tom classico inv}
\end{equation}

Since the classical probability distribution $f(p,q)$ is normalized, the
classical tomogram is also normalized for any values of the parameters $\mu $
and $\nu ,$ i.e.
\begin{equation}
\int \mathcal{W}(X,\mu ,\nu )dX=1.
\end{equation}
For $\mu =1$ , $\nu =0$ the tomogram provides the marginal distribution of
the position $(X=q)$
\begin{equation}
\mathcal{W}(X,1,0)=\int f(p,q)dp.
\end{equation}
For $\mu =0$ , $\nu =1$ the tomogram provides the marginal probability
distribution of the momentum $(X=p)$
\begin{equation}
\mathcal{W}(X,0,1)=\int f(p,q)dq.
\end{equation}

The expression (\ref{tom classico}) is written for the case of stationary
distribution function in phase space. But it is easily generalized to
time-dependent functions $f(p,q;t)$
\begin{equation}
\mathcal{W}(X,\mu ,\nu ;t)=\int f(p,q;t)\delta (X-\mu q-\nu p)dpdq.
\label{tomotime}
\end{equation}
For distribution functions associated with particle motion along a given
trajectory in phase space $q(t),p(t)$
\begin{equation}
\widetilde{f}(p,q;t)=\delta (p-p(t))\delta (q-q(t))
\end{equation}
the tomogram given by Eq.(\ref{tomotime}) reads
\begin{equation}
\widetilde{\mathcal{W}}_{f}(X,\mu ,\nu ;t)=\delta (X-\mu q(t)-\nu p(t)).
\label{tomtime1}
\end{equation}
For example, classical free motion with initial position $q_{0}$ and initial
momentum $p_{0},$ i.e. $q(t)=q_{0}+p_{0}t$;\ $p(t)=p_{0},$ is described by
the tomogram
\begin{equation}
\widetilde{\mathcal{W}}_{f}(X,\mu ,\nu ;t)=\delta (X-\mu (q_{0}+p_{0}t)-\nu
p_{0}).  \label{tomtime2}
\end{equation}

Below we will compare tomograms associated with motion along classical
trajectories with classical limits of their corresponding quantum tomogram.
When the trajectory is periodic it is possible to introduce the
time-averaged classical tomogram
\begin{equation}
\left\langle \mathcal{W}\right\rangle (X,\mu ,\nu )=\frac{1}{T}\int_{0}^{T}
\mathcal{W}(X,\mu ,\nu ;t)dt.  \label{averagedtom}
\end{equation}
The time-averaged tomogram corresponds to the time-averaged distribution
function on phase-space
\begin{equation}
\left\langle f\right\rangle (p,q)=\frac{1}{T}\int_{0}^{T}f(p,q;t)dt.
\label{rageddistribution}
\end{equation}

\subsection{Quantum tomograms}

Given a density matrix $\rho (x,x^{\prime }),$ one constructs the
corresponding Wigner function $W$ as:
\begin{equation}
W(p,q)=\int \rho (q+\frac{u}{2},q-\frac{u}{2})\exp (-i\frac{pu}{\hbar })du,
\label{1}
\end{equation}
and the Radon transform of Wigner function\thinspace $W$ is the quantum
tomogram $\mathcal{W}$ of $\rho ,$ which is a positive function\cite{Mancini}
of three variables,
\begin{equation}
\mathcal{W}(X,\mu ,\nu )=\frac{1}{2\pi }\int W(p,q)\delta (X-\mu q-\nu p)%
\frac{dpdq}{\hbar }  \label{tom quant}
\end{equation}
\begin{equation}
=\frac{1}{2\pi \hbar \left| \nu \right| }\int \rho (q+\frac{u}{2},q-\frac{u}{%
2})\exp \left[ -i\frac{X-\mu q}{\hbar \nu }u\right] dqdu  \label{tomrho}
\end{equation}
Equation (\ref{tom quant}) can be written in a form\ similar to Eq.(\ref
{tom'}),
\begin{equation}
\mathcal{W}(X,\mu ,\nu )=\left\langle \delta (X-\mu \widehat{q}-\nu \widehat{%
p})\right\rangle .  \label{tom"}
\end{equation}
The difference with Eq.(\ref{tom'}) is that here the position and momentum
are quantum operators $\hat{q}$ and $\hat{p}$ , i.e. we use non-commutative
geometry of phase-space plane, taking into account uncertainty relations.
The averaging in Eq.(\ref{tom"}) is done using density operators $\widehat{%
\rho }$, i.e.
\begin{equation}
\left\langle \widehat{A}\right\rangle :=Tr\widehat{\rho }\widehat{A}.
\end{equation}
The inverse formula of Eq.(\ref{tom quant}) is readily written as:
\begin{equation}
W(p,q)=\frac{\hbar }{2\pi }\int \mathcal{W}(X,\mu ,\nu )\exp \left[ i(X-\mu
q-\nu p)\right] dXd\mu d\nu .
\end{equation}
The density matrix $\rho (x,x^{\prime })$ can be obtained from the Wigner
function as
\begin{equation}
\rho (x,x^{\prime })=\frac{1}{2\pi \hbar }\int W(p,\frac{x+x^{\prime }}{2}%
)\exp \left[ i\frac{p(x-x^{\prime })}{\hbar }\right] dp.
\end{equation}
In terms of tomogram we get:
\begin{equation}
\rho (x,x^{\prime })=\frac{1}{2\pi }\int \mathcal{W}(X,\mu ,\frac{%
x-x^{\prime }}{\hbar })\exp \left[ i(X-\mu \frac{x+x^{\prime }}{2})\right]
dXd\mu
\end{equation}
For pure states $\rho (x,x^{\prime })=\psi (x)\psi ^{\ast }(x^{\prime })$
and $\mathcal{W}$ can be computed from Eq.(\ref{tomrho}):
\begin{equation*}
\mathcal{W}(X,\mu ,\nu )=\frac{1}{2\pi \hbar |\nu |}\int \psi (q+\frac{u}{2}%
)\psi ^{\ast }(q-\frac{u}{2})e^{-i(X-\mu q)\frac{u}{\nu \hbar }}dqdu
\end{equation*}
\begin{equation}
=\frac{1}{2\pi \hbar \left| \nu \right| }\left| \int \psi (y)e^{i\frac{\mu }{%
2\hbar \nu }y^{2}-i\frac{X}{\hbar \nu }y}dy\right| ^{2},\quad \nu \neq 0.
\label{tom quant pure state 1}
\end{equation}
Thus, apart from the pre-factor $1/2\pi \hbar \left| \nu \right| $, the
tomogram $\mathcal{W}$ of the wave function $\psi $ is the square modulus of
the tomogram amplitude $A_{\psi }:$%
\begin{equation}
A_{\psi }(X,\mu ,\nu ):=\int \psi (y)e^{i\frac{\mu }{2\hbar \nu }y^{2}-i%
\frac{X}{\hbar \nu }y}dy.
\end{equation}
When $\nu =0$ and $\mu \neq 0$ one can instead use
\begin{equation}
\mathcal{W}(X,\mu ,\nu )=\frac{1}{2\pi \hbar \left| \mu \right| }\left| \int
\widehat{\psi }(p)e^{-i\frac{\nu }{2\hbar \mu }p^{2}+i\frac{X}{\hbar \mu }%
p}dp\right| ^{2},\quad \mu \neq 0,  \label{tom quant pure state 2}
\end{equation}
where the Fourier transform of the wave function has been introduced
\begin{equation}
\widehat{\psi }(p)=\frac{1}{\sqrt{2\pi \hbar }}\int \psi (y)e^{-i\frac{py}{%
\hbar }}dy.  \label{fourier}
\end{equation}
If both $\mu =\nu =0$, we get $\mathcal{W}(X,0,0)=\delta (X)$, after using $%
\int \rho (x,x)dx=1$.

Tomograms of quantum states and classical statistical density are positive
functions on the same space $\mathbb{R}^{3}$ and therefore can be compared.
Moreover, the same dimensional relations hold in both cases:
\begin{equation}
\lbrack X]=[\mu ][q]=[\nu ][p]\quad ;\quad \lbrack \mathcal{W}(X,\mu ,\nu
)]=[X]^{-1}
\end{equation}
where $[.]$ indicates the units. $X$ and $\mathcal{W}$ can be made
dimensionless after assuming
\begin{equation}
\lbrack \mu ]=[q]^{-1}\quad ;\quad \lbrack \nu ]=[p]^{-1}.
\end{equation}

The operator
\begin{equation}
\widehat{X}=\mu \widehat{q}+\nu \widehat{p}
\end{equation}
together with its conjugate
\begin{equation}
\widehat{Y}=-s^{2}\nu \widehat{q}+s^{-2}\mu \widehat{p}
\end{equation}
preserves the canonical commutation relations: $[\widehat{X},\widehat{Y}]=[%
\widehat{q},\widehat{p}].$ The observable $\widehat{X}$ is a new position
operator, i.e. the position after a symplectic (linear canonical)
transformation in the quantum phase-space $(\widehat{q},\widehat{p})$ of the
particle. The real variable $X$ gives the possible results of a measure of$\
\widehat{X}$ and runs over the spectrum of $\widehat{X}$. In this way a
description of quantum tomograms is recovered in complete analogy with the
classical case. So the tomogram is also ``symplectic'' in the quantum case.

In classical mechanics, the transition from the distribution function of two
canonically conjugate variables (position $q$ and momentum $p$) to the
distribution function of one variable (position $X$) does not play a crucial
role due to absence of quantum mechanical constraints like the uncertainty
relations of Heisenberg\cite{Heis27} and Robertson-Schroedinger\cite
{MarSud....,Rob29,Shr30,Dod,SudBh}. On the contrary, the use of tomograms in
quantum mechanics provides the possibility of describing a quantum state
with a probability distribution which does not violate the uncertainty
relations. Comparing the Eqs.(\ref{tom'}) and (\ref{tom"}) we see that the
classical limit will be found starting from a non-commutative plane $(%
\widehat{q},\widehat{p})$ and going to a commutative one $(q,p).$ This means
that in the limit process we take into full account the behaviour of the
quantum uncertainty relations.

\section{The quantum-classical transition}

The classical limit of quantum states can be obtained using different
procedures. One procedure is to take the limit $\hbar \rightarrow 0$ of
certain quantum mechanical expressions. The wave function defined on the
configuration space is not suitable for this purpose, since there is no
physical interpretation of the resulting limit. It is more useful to
associate an "observable" with any state vector, namely the associated
rank-one projector. Following the Weyl-Wigner approach observables can be
mapped to functions on phase-space and finally to tomograms via Radon
transform. The classical limit $\hbar \rightarrow 0$ then can be done on
tomograms in two different ways.

\begin{enumerate}
\item  At a kinematic level (Planck limit). Here no reference is made to any
specific dynamics.

\item  At a dynamical level (Ehrenfest limit). Here the limit $\hbar
\rightarrow 0$ on tomograms is performed keeping constant the mean value of
a given observable (usually the Hamiltonian).
\end{enumerate}

\subsection{Planck limits}

We can anticipate some general features of the Planck limit for the tomogram
of a quantum state. The quantum fluctuations of position and momentum,
depending on Planck's constant, are washed out when $\hbar \rightarrow 0$
and in general a Dirac $\delta (X)$ results, corresponding to the tomogram
of the classical probability distribution $\delta (p)\delta (q).$ The fact
that that Dirac delta function is centered at the origin has no particular
physical significance, as it is only due to the arbitrary choice of the
origin of the affine phase space.

The Planck limit can be predicted from a scaling argument for a large class
of wave functions. Assume the dependence on $\hslash $ in the wave function $%
\psi $ is ($\gamma $ a real number)
\begin{equation}
\psi (x)=\hbar ^{\frac{\gamma }{2}}\Psi (\hbar ^{\gamma }x).
\end{equation}
This assures that
\begin{equation}
\int |\psi (x)|^{2}dx=1
\end{equation}
independent of the value of $\hbar $\textit{.} Computing the tomogram in
this case we get
\begin{eqnarray}
\mathcal{W}(X,\mu ,\nu ) &=&\frac{1}{2\pi \hbar \left| \nu \right| }\left|
\int \Psi (\hbar ^{\gamma }y)\hbar ^{\frac{\gamma }{2}}e^{i\frac{\mu }{2\nu
\hbar }y^{2}-i\frac{X}{\nu \hbar }y}dy\right| ^{2}  \notag \\
&=&\frac{1}{2\pi \left| \nu \right| \hbar ^{\gamma +1}}\left| \int \Psi
(t)e^{i\frac{\mu }{2\nu \hbar ^{2\gamma +1}}t^{2}-i\frac{X}{\nu \hbar
^{\gamma +1}}t}dt\right| ^{2}
\end{eqnarray}

When $\gamma =-1/2$ it gives
\begin{equation}
\mathcal{W}(X,\mu ,\nu )=\frac{1}{2\pi \left| \nu \right| \sqrt{\hbar }}%
\left| \int \Psi (t)e^{i\frac{\mu }{2\nu }t^{2}-i\frac{X}{\nu \sqrt{\hbar }}%
t}dt\right| ^{2}=\frac{1}{\sqrt{\hbar }}\mathcal{F}(\frac{X}{\sqrt{\hbar }}%
,\mu ,\nu ).
\end{equation}
Since in general $\int \mathcal{W}(X,\mu ,\nu )dX=1,$ the Dirac delta
theorem of the Appendix, with $n=1/\sqrt{\hbar },$ applies and yields the
Planck limit of the tomogram as
\begin{equation}
\lim_{\hbar \rightarrow 0}\mathcal{W}(X,\mu ,\nu )=\delta (X).
\label{Planck}
\end{equation}
This is the case of the harmonic oscillator eigenstates and their
superpositions.

When $\gamma =0$ for the wave function in position space, its Fourier
transform $\widehat{\psi }$ of Eq.(\ref{fourier}) scales with $\gamma =-1/2$
and using Eq.(\ref{tom quant pure state 2}) we recover the same Planck limit
of Eq.(\ref{Planck}). In general this happens when a quantum particle of
mass $m$ is confined in a potential with a characteristic length $L,$ then $%
V_{0}=\hbar ^{2}/2mL^{2}$ is the scale factor of the potential energy so
that the eigenvalues are proportional to $V_{0}$ while the eigenstates do
not depend on $\hbar $. This is the case of the Poeschl-Teller potential\cite
{poe-teller} and of its limit, the infinite square well.

We finally observe that a more general scaling law could be
\begin{equation}
\psi (x)=\hbar ^{\frac{\gamma }{2}}\Psi (\hbar ^{\gamma }(x-x_{0})).
\end{equation}
In that case, when $\gamma =-1/2,$ the Planck limit of the tomogram is $%
\delta (X-\mu x_{0})$ and corresponds to the classical distribution $\delta
(p)\delta (q-x_{0}).$ Nevertheless, by using the shifted position operator $%
\widehat{q}-x_{0},$ the Planck limit turns out to be again $\delta (X).$

In the following we study the Planck limit for examples involving Hermite
eigenfunctions. We start by reviewing the Hermite tomograms, whose
expressions will also be useful in evaluating the Ehrenfest limit.

\subsubsection{Harmonic oscillator}

Recall that the eigenstates of a quantum harmonic oscillator of mass $m$ and
frequency $\omega ,$ normalized with respect to the Lebesgue measure $dy,$
are
\begin{equation}
\varphi _{n}(\sqrt{\frac{\varpi }{\hslash }}y)=(\sqrt{\pi }2^{n}n!)^{-\frac{1%
}{2}}\sqrt[4]{\frac{\varpi }{\hslash }}H_{n}(\sqrt{\frac{\varpi }{\hslash }}%
y)\exp \left[ -\frac{\varpi }{2\hslash }y^{2}\right]
\end{equation}
with $\varpi =m\omega $ and $n=0,1,2,...$ Using Eq.(\ref{tom quant pure
state 1}) the tomogram $\mathcal{W}_{0}(X,\mu ,\nu )$ for the ground state
can be computed as
\begin{equation}
\frac{1}{\sqrt{\pi }}\sqrt{\frac{\varpi }{\hslash (\varpi ^{2}\nu ^{2}+\mu
^{2})}}\exp \left[ -\frac{\varpi }{\hslash (\varpi ^{2}\nu ^{2}+\mu ^{2})}%
X^{2}\right] =\varphi _{0}^{2}\left( \sqrt{\frac{\varpi }{\hslash (\varpi
^{2}\nu ^{2}+\mu ^{2})}}X\right) .
\end{equation}
Similarly, the tomogram $\mathcal{W}_{n}$ for the $n$-th excited state is
\begin{equation}
\mathcal{W}_{n}(X,\mu ,\nu )=\varphi _{n}^{2}\left( \sqrt{\frac{\varpi }{%
\hslash (\varpi ^{2}\nu ^{2}+\mu ^{2})}}X\right)  \label{HermTom}
\end{equation}
A suitable way to obtain these results is to evaluate a generating function $%
J$ for the tomogram amplitudes $A_{n}$ of Hermite functions $\varphi _{n}$
\begin{equation*}
J(s):=\sum\limits_{n=0}^{\infty }\frac{s^{n}}{\sqrt{n!}}A_{n}=\sum%
\limits_{n=0}^{\infty }\frac{s^{n}}{\sqrt{n!}}\int \varphi _{n}(\sqrt{\frac{%
\varpi }{\hslash }}y)\exp \left[ i\frac{\mu }{2\hbar \nu }y^{2}-i\frac{X}{%
\hbar \nu }y\right] dy
\end{equation*}
\begin{equation}
=\sqrt[4]{\frac{\varpi }{\pi \hbar }}\sqrt{\frac{2\pi \hbar \nu }{\zeta
^{\ast }}}\exp \left[ \frac{\zeta }{2\zeta ^{\ast }}s^{2}-i\sqrt{\frac{%
2\varpi }{\hbar }}\frac{Xs}{\zeta ^{\ast }}-\frac{1}{2\hbar \nu \zeta ^{\ast
}}X^{2}\right]  \label{generating}
\end{equation}
where $\zeta =\varpi \nu +i\mu $ and $\zeta ^{\ast }$ is its complex
conjugate. It is possible to reconstruct a generating function of Hermite
functions $\varphi _{n}.$ Defining
\begin{equation}
\tau =-i\sqrt{\frac{\zeta }{\zeta ^{\ast }}}s\quad ;\quad Q=\sqrt{\frac{%
\varpi }{\hbar \zeta \zeta ^{\ast }}}X,
\end{equation}
Equation (\ref{generating}) becomes
\begin{equation}
J(\tau )=\sqrt{2\pi \hbar \nu }\sqrt[4]{\frac{\zeta }{\zeta ^{\ast }}}\exp %
\left[ -i\frac{\mu }{2\varpi \nu }Q^{2}\right] \sum\limits_{n=0}^{\infty }%
\frac{\tau ^{n}}{\sqrt{n!}}\varphi _{n}(Q).
\end{equation}
Eventually, the Hermite tomogram amplitudes can be written
\begin{equation}
A_{n}=\frac{1}{\sqrt{n!}}\frac{d^{n}J}{ds^{n}}(0)=\sqrt{2\pi \hbar \nu }\exp %
\left[ -i\frac{\mu }{2\varpi \nu }Q^{2}\right] \sqrt[4]{\frac{\zeta }{\zeta
^{\ast }}}\left( -i\sqrt{\frac{\zeta }{\zeta ^{\ast }}}\right) ^{n}\varphi
_{n}(Q),  \label{HermiteAmp}
\end{equation}
which yield Hermite tomograms $\mathcal{W}_{n}$\ of Eq.(\ref{HermTom}) for $%
\varphi _{n}.$ Observing that $\varphi _{n}$ is normalized to unity, the
Planck limit of $\mathcal{W}_{n},$ evaluated by means of the Dirac delta
theorem (see Appendix), is
\begin{equation}
\underset{\hbar \rightarrow 0}{\lim }\mathcal{W}_{n}(X,\mu ,\nu )=\delta (X)
\label{HermitePlanck}
\end{equation}
as expected.

\subsubsection{Superposition}

The case of a superposition of two harmonic oscillator eigenstates
\begin{equation}
\psi =\frac{1}{\sqrt{2}}\left( \varphi _{n}+\varphi _{m}\right)
\end{equation}
shows in the simplest manner the vanishing of quantum interference in the
classical Planck limit. Writing the tomogram in terms of Hermite amplitudes
of Eq.(\ref{HermiteAmp})
\begin{equation}
\mathcal{W}_{\psi }(X,\mu ,\nu )=\frac{1}{2\pi \hbar \left| \nu \right| }%
\left| \frac{1}{\sqrt{2}}\left( A_{n}+A_{m}\right) \right| ^{2}=\frac{1}{2}%
\mathcal{\ W}_{n}+\frac{1}{2}\mathcal{W}_{m}+\mathop{\mathrm{Re}}\frac{%
A_{n}A_{m}^{\ast }}{2\pi \hbar \left| \nu \right| }.
\end{equation}
Dropping a unimodular factor independent of $\hbar $, the quantum
interference term reads
\begin{equation}
\frac{A_{n}A_{m}^{\ast }}{2\pi \hbar \left| \nu \right| }\varpropto \varphi
_{n}(Q)\varphi _{m}(Q).
\end{equation}
Due to the orthogonality of harmonic oscillator eigenstates, this term
vanishes when $\hbar \rightarrow 0$ as is shown in the Appendix. Moreover,
it is possible to show that the quantum interference term vanishes like $%
\sqrt{\hbar }$. Again, by using Eq.(\ref{HermitePlanck}), the classical
Planck limit of superposition state tomogram is
\begin{equation}
\underset{\hbar \rightarrow 0}{\lim }\mathcal{W}_{\psi }(X,\mu ,\nu )=\delta
(X).
\end{equation}

\subsubsection{Coherent states}

We next consider the tomogram of coherent states. Recalling that the
coherent state $\left| \alpha \right\rangle $ labeled by the complex number $%
\alpha $ is defined by
\begin{equation}
\left| \alpha \right\rangle :=\mathrm{e}^{-\left| \alpha \right|
^{2}/2}\sum\limits_{n=0}^{\infty }\frac{\alpha ^{n}}{\sqrt{n!}}\left|
n\right\rangle ,
\end{equation}
the tomogram amplitude $A_{\alpha }$ of the coherent state can be expressed
through Hermite amplitudes of Eq.(\ref{HermiteAmp})
\begin{equation}
A_{\alpha }=\mathrm{e}^{-\left| \alpha \right|
^{2}/2}\sum\limits_{n=0}^{\infty }\frac{\alpha ^{n}}{\sqrt{n!}}A_{n}=\mathrm{%
e}^{-\left| \alpha \right| ^{2}/2}J(\alpha )
\end{equation}
From the expression for the generating function of Hermite amplitudes $J$ in
Eq.(\ref{generating}), the previous equation becomes
\begin{equation}
A_{\alpha }=\sqrt[4]{\frac{\varpi }{\pi \hbar }}\sqrt{\frac{2\pi \hbar \nu }{%
\zeta ^{\ast }}}\mathrm{e}^{-i\frac{\mu }{2\varpi \nu }Q^{2}}\mathrm{e}^{-%
\frac{\left| \alpha \right| ^{2}}{2}}\exp \left[ -\frac{Q^{2}}{2}-i\alpha
\sqrt{2\frac{\zeta }{\zeta ^{\ast }}}Q+\frac{\zeta \alpha ^{2}}{2\zeta
^{\ast }}\right]  \label{ampcohe}
\end{equation}
so that, eventually, the tomogram $\mathcal{W}_{\alpha }(X,\mu ,\nu
)=A_{\alpha }A_{\alpha }^{\ast }/2\pi \hbar \left| \nu \right| $ is given by
\begin{equation}
\sqrt{\frac{\varpi }{\pi \hbar (\varpi ^{2}\nu ^{2}+\mu ^{2})}}\exp \left[ -%
\frac{\left( \sqrt{\varpi }X-\mu \sqrt{2\hbar }\mathop{\mathrm{Re}}\alpha
-\varpi \nu \sqrt{2\hbar }\mathop{\mathrm{Im}}\alpha \right) ^{2}}{\hbar
(\varpi ^{2}\nu ^{2}+\mu ^{2})}\right] .  \label{cohetomo}
\end{equation}
After shifting the integration variable
\begin{equation}
\int \mathcal{W}_{\alpha }(X,\mu ,\nu )dX=\int \frac{1}{\sqrt{\pi }}\exp %
\left[ -\xi ^{2}\right] d\xi =1,
\end{equation}
and by the Dirac delta theorem the Planck limit of the tomogram is
\begin{equation}
\mathcal{W}_{\alpha }(X,\mu ,\nu )\underset{\hbar \rightarrow 0}{%
\longrightarrow }\delta (X).
\end{equation}

\subsubsection{Schroedinger cat states}

Even and odd coherent states (Schroedinger cat states) are defined\cite
{Phisica74} as:
\begin{equation}
\left| \alpha _{\pm }\right\rangle =N_{\pm }(\left| \alpha \right\rangle \pm
\left| -\alpha \right\rangle )
\end{equation}
where
\begin{equation}
\left\langle \beta |\alpha \right\rangle =\exp \left[ -\frac{\left| \alpha
\right| ^{2}}{2}-\frac{\left| \beta \right| ^{2}}{2}+\alpha \beta ^{\ast }%
\right] \Rightarrow N_{\pm }=\frac{1}{\sqrt{2(1\pm e^{-2|\alpha |^{2}})}}.
\end{equation}
The tomograms $\mathcal{W}_{\alpha _{\pm }}(X,\mu ,\nu )=N_{{\pm }%
}^{2}\left| A_{\alpha }\pm A_{-\alpha }\right| ^{2}/2\pi \hbar \left| \nu
\right| $ contain three terms
\begin{equation}
N_{{\pm }}^{2}\left\{ \mathcal{W}_{\alpha }(X,\mu ,\nu )+\mathcal{W}%
_{-\alpha }(X,\mu ,\nu )\pm I(X,\mu ,\nu )\right\} .  \label{cattomo}
\end{equation}
The interference term reads
\begin{equation}
I(X,\mu ,\nu )=2\mathop{\mathrm{Re}}\frac{A_{\alpha }A_{-\alpha }^{\ast }}{%
2\pi \hbar \left| \nu \right| }
\end{equation}
with tomogram amplitude $A_{\alpha }$ given by Eq.(\ref{ampcohe}). As shown
in the Remark of Appendix, it results
\begin{equation}
\int \frac{A_{\alpha }A_{-\alpha }^{\ast }}{2\pi \hbar \left| \nu \right| }%
dX=\left\langle -\alpha |\alpha \right\rangle =e^{-2\left| \alpha \right|
^{2}}
\end{equation}
and by Dirac delta theorem the Planck limit of the interference term $%
I(X,\mu ,\nu )$ is $2e^{-2\left| \alpha \right| ^{2}}\delta (X).$ Then the
Planck limit of the Schroedinger cat tomogram is
\begin{equation}
\underset{\hbar \rightarrow 0}{\lim }\mathcal{W}_{\alpha _{\pm }}(X,\mu ,\nu
)=N_{{\pm }}^{2}(2\delta (X)\pm 2e^{-2\left| \alpha \right| ^{2}}\delta
(X))=\delta (X).
\end{equation}
We observe that the Planck limit $\delta (X)$ is gained thanks to the
contribution of a non-zero interference term between non-orthogonal states.

\subsection{Ehrenfest classical limits}

The Ehrenfest classical limit takes place at the dynamical level, i.e. it
requires a specific dynamical system and the limit is performed while
keeping constant the value of some selected observable, usually the energy.
\cite{MarSud....} In this section some Ehrenfest classical limits are
evaluated with the constraint of constant energy.

We observe that after calculating the Ehrenfest limit, it is possible to
consider the limit when the energy goes to zero. This is not equivalent to
the Planck limit because in the Ehrenfest case, if there is a potential with
a minimum, the particle goes to its equilibrium position $q_{0}$ when the
energy vanishes. Then the expected result is $\delta (X-\mu q_{0})$, rather
than $\delta (X)$, agreeing with the classical tomogram of a rest state $%
\delta (p)\delta (X-q_{0})$, as the next example of the coherent state shows.

\subsubsection{Schroedinger cats}

The Ehrenfest classical limit of the coherent state tomogram $\mathcal{W}%
_{\alpha }$ is readily evaluated from Eq.(\ref{cohetomo}). Put $\varpi =1$
and\ impose the constraint on the energy by taking constant the mean value
of the harmonic oscillator Hamiltonian $\hbar (\widehat{a}^{\dagger }%
\widehat{a}+1/2)$ on the coherent state $\left| \alpha \right\rangle $. The
constraint is satisfied by choosing
\begin{equation}
\alpha \sqrt{\hbar }=const\Rightarrow \hbar \alpha ^{\ast }\alpha =\hbar
\left\langle \widehat{a}^{\dagger }\widehat{a}\right\rangle _{\alpha
}=const\ .
\end{equation}
Now, bearing in mind that the mean value of the annihilation operator $%
\widehat{a}$ on the coherent state $\left| \alpha \right\rangle $ is
\begin{equation}
\left\langle \widehat{a}\right\rangle _{\alpha }=\frac{\left\langle \widehat{%
q}+i\widehat{p}\right\rangle _{\alpha }}{\sqrt{2\hbar }}=\frac{q_{\alpha
}+ip_{\alpha }}{\sqrt{2\hbar }}=\alpha \ ,
\end{equation}
the Ehrenfest classical limit $\hbar \rightarrow 0$ , $\alpha \rightarrow
\infty $ with\ $\alpha \sqrt{2\hbar }=const=q_{\alpha }+ip_{\alpha }$ turns
out to be, just by inspection of Eq.(\ref{cohetomo}) (with $\varpi =1$), the
displaced delta function
\begin{equation}
\delta (X-\mu q_{\alpha }-\nu p_{\alpha })
\end{equation}
which coincides with the tomogram of the classical distribution function
\begin{equation}
f(p,q)=\delta (p-p_{\alpha })\delta (q-q_{\alpha }).
\end{equation}

As far as the Schroedinger cat states of Eq.(\ref{cattomo}) are concerned,
the interference term $I(X,\mu ,\nu )$ vanishes in the Ehrenfest limit, due
to the presence of a rapidly oscillating factor ($\hbar \rightarrow 0)$%
\begin{equation}
\cos \frac{2\sqrt{2\hbar }X(\nu \mathop{\mathrm{Re}}\alpha -\mu %
\mathop{\mathrm{Im}}\alpha )}{\hbar (\nu ^{2}+\mu ^{2})}=\cos \frac{2X(\nu
q_{\alpha }-\mu p_{\alpha })}{\hbar (\nu ^{2}+\mu ^{2})}
\end{equation}
while the normalizations $N_{\pm }^{2}$ go to the same constant value
\begin{equation}
N_{\pm }^{2}=\frac{1}{2(1\pm e^{-2|\alpha |^{2}})}\underset{\alpha
\rightarrow \infty }{\longrightarrow }\frac{1}{2}
\end{equation}
\ and the tomograms become the same sum of delta functions
\begin{equation}
\mathcal{W}_{\infty }(X,\mu ,\nu )=\frac{1}{2}[\delta (X-\mu q_{\alpha }-\nu
p_{\alpha })+\delta (X+\mu q_{\alpha }+\nu p_{\alpha })].
\end{equation}

\subsubsection{Infinite square well}

We consider a particle of mass $m=1$ in an infinite square well $0\leq q\leq
L.$ To deal with wave functions of compact support it is convenient to
regard them as elements of $L_{2}(\mathbb{R})$ suitably projected by means
of the projector $\chi _{\lbrack 0,L]}$ , which is the standard
characteristic function of the set $[0,L]$.\ The eigenfunctions and the
eigenvalues are
\begin{equation}
\psi _{n}(q)=\sqrt{\frac{2}{L}}\sin (\frac{n\pi }{L}q)\chi _{\lbrack
0,L]}(q)\ ;\ E_{n}=\frac{1}{2}(\frac{\hbar n\pi }{L})^{2},\quad
n=1,2,...\quad .
\end{equation}
Fixing the energy to be one for convenience, the Ehrenfest classical limit
amounts to
\begin{equation}
\hbar =\frac{\sqrt{2}L}{n\pi },\quad n\rightarrow \infty .
\end{equation}
In general, for $\nu \neq 0,$
\begin{equation}
\mathcal{W}(X,\mu ,\nu )=\frac{1}{L\pi \hbar \left| \nu \right| }\left|
\int\nolimits_{0}^{L}\sin (\frac{n\pi }{L}y)e^{i\frac{\mu }{2\hbar \nu }%
y^{2}-i\frac{X}{\hbar \nu }y}dy\right| ^{2}=\frac{n\left| A_{-}-A_{+}\right|
^{2}}{4L^{2}\sqrt{2}\left| \nu \right| }  \label{tomwe}
\end{equation}
where the amplitudes $A_{\mp }$ are respectively given by
\begin{equation}
\int\nolimits_{0}^{L}\exp \left[ inF_{\mp }(y)\right] dy:=\int%
\nolimits_{0}^{L}\exp \left\{ in\frac{\pi }{L}\left[ \frac{\mu }{2\sqrt{2}%
\nu }y^{2}-(\frac{X}{\sqrt{2}\nu }\mp 1)y\right] \right\} dy.
\label{tomwell}
\end{equation}
These amplitudes can be computed exactly in terms of Error function.
However, in view of the large $n$ limit, it is more useful to evaluate these
integrals in the stationary phase approximation. It results
\begin{equation}
A_{\mp }\simeq \sqrt{\frac{2L\left| \nu \right| \sqrt{2}}{n\left| \mu
\right| }}\exp \left\{ i\left[ nF_{\mp }(Q_{s}^{\mp })+\mathrm{sign\ }(\frac{%
\mu }{\nu })\frac{\pi }{4}\right] \right\} \ \chi _{\lbrack 0,L]}(Q_{s}^{\mp
}).
\end{equation}
where the stationary points are
\begin{equation}
Q_{s}^{\mp }=\frac{X}{\mu }\mp \frac{\sqrt{2}\nu }{\mu }.  \label{Qstaz}
\end{equation}
Eventually, the asymptotic behaviour of $\mathcal{W}(X,\mu ,\nu )$ is
\begin{equation*}
\frac{ \chi _{\lbrack 0,L]}(Q_{s}^{-})+\chi _{\lbrack 0,L]}(Q_{s}^{+})-2\chi
_{\lbrack 0,L]}(Q_{s}^{-})\chi _{\lbrack 0,L]}(Q_{s}^{+})\cos
n[F_{-}(Q_{s}^{-})-F_{+}(Q_{s}^{+})]}{2\left| \mu \right| L}
\end{equation*}
and evaluating the Ehrenfest classical limit $n\rightarrow \infty $ in the
sense of a distribution:
\begin{equation}
\mathcal{W}_{\infty }(X,\mu ,\nu )=\frac{1}{2\left| \mu \right| L}\left\{
\chi _{\lbrack 0,L]}(Q_{s}^{-})+\chi _{\lbrack 0,L]}(Q_{s}^{+})\right\} .
\end{equation}
This result has to be compared with its classical analogue. In that case,
bearing in mind that $m=1$ and $p=\pm \sqrt{2E}=\pm \sqrt{2},$ we have
\begin{equation}
f(p,q;t)=\delta (p-p(t))\delta (q-q(t))
\end{equation}
where $q(t)$ is a zigzag line of height $L$ and half-period $L/\sqrt{2}$,
\begin{equation}
q(t)=\sqrt{2}t\chi _{\lbrack 0,L/\sqrt{2}]}(t)-\sqrt{2(}t-\sqrt{2}L)\chi
_{\lbrack L/\sqrt{2},\sqrt{2}L]}(t),
\end{equation}
while
\begin{equation}
p(t)=\sqrt{2}\chi _{\lbrack 0,L/\sqrt{2}]}(t)-\sqrt{2}\chi _{\lbrack L/\sqrt{%
2},\sqrt{2}L]}(t).
\end{equation}

To make a comparison with the Ehrenfest limit we have to average the
classical time dependent distribution tomogram over a period:
\begin{equation*}
\mathcal{W}(X,\mu ,\nu )=\frac{1}{\sqrt{2}L}\int\nolimits_{0}^{\sqrt{2}%
L}dt\int f(p,q;t)\delta (X-\mu q-\nu p)dpdq
\end{equation*}
\begin{equation}
=\frac{1}{\sqrt{2}L\left| \mu \right| }\left( \int\nolimits_{0}^{\sqrt{2}%
L/2}\delta (Q_{s}^{-}-q(t))dt+\int\nolimits_{\sqrt{2}L/2}^{\sqrt{2}L}\delta
(Q_{s}^{+}-q(t))dt\right) ,
\end{equation}
where $Q_{s}^{\mp }$ are again given by Eq.(\ref{Qstaz}). Eventually,
\begin{equation}
\mathcal{W}(X,\mu ,\nu )=\frac{1}{2\left| \mu \right| L}\left\{ \chi
_{\lbrack 0,L]}(Q_{s}^{-})+\chi _{\lbrack 0,L]}(Q_{s}^{+})\right\} =\mathcal{%
\ W}_{\infty }(X,\mu ,\nu ).
\end{equation}
This general result yields the marginal probability distribution of $q,$ in the limit $%
\nu \rightarrow 0, \mu =1$
\begin{equation}
\underset{\nu \rightarrow 0,\mu \rightarrow 1}{\lim }\frac{1}{2\left| \mu
\right| L}\left\{ \chi _{\lbrack 0,L]}(Q_{s}^{-})+\chi _{\lbrack
0,L]}(Q_{s}^{+})\right\} =\frac{1}{L}\chi _{\lbrack 0,L]}(X).
\end{equation}
As expected, this result shows that the particle position is likely
distributed in the box.

The limit $\nu \rightarrow 1, \mu =0$ which yields the marginal
probability distribution of $p$ is more involved. However, observing
that
\begin{equation}
\int \frac{1}{\left| \mu \right| L}\chi _{\lbrack 0,L]}(Q_{s}^{\mp })dX=1
\end{equation}
and remembering the explicit expression of $Q_{s}^{\mp },$\ by Dirac delta
theorem it results
\begin{equation}
\underset{\nu \rightarrow 1,\mu \rightarrow 0}{\lim }\frac{1}{2\left| \mu
\right| L}\left\{ \chi _{\lbrack 0,L]}(Q_{s}^{-})+\chi _{\lbrack
0,L]}(Q_{s}^{+})\right\} =\frac{1}{2}\delta (X-\sqrt{2})+\frac{1}{2}\delta
(X+\sqrt{2})
\end{equation}
As expected, the probability distribution of momentum is likely concentrated
on the two allowed classical values of $p.$

\subsubsection{Harmonic oscillator}

We assume $m=1,\omega =1$ and evaluate the Ehrenfest classical limit of the
Hermite tomogram $\mathcal{W}_{n}$ of Eq.(\ref{HermTom}) when
\begin{equation}
\hbar \rightarrow 0,E_{n}=const=1\Rightarrow \left( n+\frac{1}{2}\right)
\hbar =1.
\end{equation}
In view of the large $n$ limit, we may assume $n+\frac{1}{2}\simeq n,$ so
that $\hbar =1/n.$ Besides, due to the spherical symmetry on the phase
space, it is enough to consider the case $\mu =1,\nu =0.$ Then the tomogram
of the harmonic oscillator eigenstate $\varphi _{n}$ is written as
\begin{equation}
\mathcal{W}_{n}(X,1,0)=\sqrt{\frac{n}{\pi }}\frac{e^{-nX^{2}}}{2^{n}n!}%
H_{n}^{2}(\sqrt{n}X)
\end{equation}
Note that the tomogram is an even function of $X$, so we only need study
positive values of $X.$ We start by expressing Hermite polynomials $H_{n}$
in terms of parabolic cylinder functions $U$ (see \cite{Abramowitz-Stegun})
\begin{equation}
H_{n}(y)=2^{n/2}e^{y^{2}/2}U(-(n+\frac{1}{2}),\sqrt{2}y).
\end{equation}
The main asymptotic formula is:
\begin{equation}
U(a,x)\simeq 2^{-1/4-a/2}\Gamma (\frac{1}{4}-\frac{a}{2})\sqrt[4]{\frac{\tau
}{\xi ^{2}-1}}Ai(\tau ),\quad \left( x\geq 0,a\rightarrow -\infty \right) ,
\end{equation}
where the Airy function $Ai$ and the Euler gamma function $\Gamma $ appear,
with:
\begin{equation}
\xi =\frac{x}{2\sqrt{\left| a\right| }};\quad \tau =(4\left| a\right| )^{2/3}%
\left[ \mp \left( \frac{3}{2}\Theta _{\lessgtr }\right) \right] ^{2/3}
\label{asint1}
\end{equation}
and correspondingly:
\begin{equation}
\Theta _{\lessgtr }=\left\{
\begin{array}{c}
\frac{1}{4}\left[ \arccos \xi -\xi \sqrt{1-\xi ^{2}}\right] ,\quad (\xi \leq
1) \\
\frac{1}{4}\left[ \xi \sqrt{\xi ^{2}-1}-\cosh ^{-1}\xi \right] ,\quad (\xi
\geq 1)
\end{array}
\right. .  \label{asint2}
\end{equation}
In our case, the asymptotic behaviour of $\mathcal{W}_{n}(X,1,0)$ is
\begin{equation}
\sqrt{\frac{n}{\pi }}\frac{1}{n!}U^{2}(-(n+\frac{1}{2}),\sqrt{2n}X)\underset{%
n\gg 1}{\simeq }\sqrt{\frac{n}{\pi }}\frac{1}{n!}2^{n}\Gamma ^{2}(\frac{n+1}{%
2})\sqrt{\frac{\tau }{\xi ^{2}-1}}Ai^{2}(\tau )
\end{equation}
Now, the numerical pre-factor can be readily estimated as:
\begin{equation}
\sqrt{\frac{n}{\pi }}\frac{1}{n!}2^{n}\Gamma ^{2}(\frac{n+1}{2})=\sqrt{n}%
\frac{\Gamma (\frac{n+1}{2})}{\Gamma (\frac{n}{2}+1)}\underset{n\rightarrow
\infty }{\simeq }\sqrt{n}\sqrt{\frac{2}{n}}=\sqrt{2}
\end{equation}
Using definitions in equations $\left( \ref{asint1}\right) $, $\left( \ref
{asint2}\right) $ we have
\begin{equation*}
\xi =\frac{\sqrt{2n}}{2\sqrt{n+\frac{1}{2}}}X\underset{n\rightarrow \infty }{%
\longrightarrow }\frac{X}{\sqrt{2}};\quad \xi \lessgtr 1\Rightarrow \tau
=\mp \left| \tau \right| ,\quad \frac{2}{3}\left| \tau \right| ^{3/2}=4(n+%
\frac{1}{2})\Theta _{\lessgtr }
\end{equation*}
and the asymptotic behaviour of the Airy function\cite{Abramowitz-Stegun}:
\begin{equation}
\left\{
\begin{array}{c}
Ai(-\left| \tau \right| )\simeq \pi ^{-1/2}\left| \tau \right| ^{-1/4}\sin
\left[ \frac{2}{3}\left| \tau \right| ^{3/2}+\frac{\pi }{4}\right]  \\
Ai(\left| \tau \right| )\simeq (4\pi )^{-1/2}\left| \tau \right| ^{-1/4}\exp
\left[ -\frac{2}{3}\left| \tau \right| ^{3/2}\right]
\end{array}
\right.
\end{equation}
we obtain the Ehrenfest classical distribution limit of $\mathcal{W}%
_{n}(X,1,0)$ :
\begin{equation*}
\frac{\sqrt{2}}{\pi \sqrt{\left| 1-\xi ^{2}\right| }}\left\{
\begin{array}{c}
\sin ^{2}\left[ (n+\frac{1}{2})(\arccos \xi -\xi \sqrt{1-\xi ^{2}})+\frac{%
\pi }{4}\right] \rightarrow \frac{1}{2}\quad (\xi \leq 1) \\
\frac{1}{4}\exp \left[ -2(n+\frac{1}{2})\left( \xi \sqrt{\xi ^{2}-1}-\cosh
^{-1}\xi \right) \right] \rightarrow 0\quad (\xi \geq 1)
\end{array}
\right.
\end{equation*}
with $\xi =X/\sqrt{2}.$ Eventually, restoring the negative values of $X,$ we
have
\begin{equation}
\mathcal{W}_{\infty }(X,1,0)=\frac{\sqrt{2}}{2\pi \sqrt{\left| 1-\left(
\frac{X}{\sqrt{2}}\right) ^{2}\right| }}\chi _{\lbrack -1,1]}(\frac{X}{\sqrt{%
2}}).  \label{Wosc_infin}
\end{equation}

The classical tomogram $\mathcal{W}(X,1,0)$ is the time average of the Radon
transform of $f(p,q;t):$%
\begin{equation}
\frac{1}{T}\int\nolimits_{0}^{T}dt\int \delta (p-p(t))\delta (q-q(t))\delta
(X-q)dpdq=\frac{1}{T}\int\nolimits_{0}^{T}\delta (X-q(t))dt
\end{equation}
where $T=2\pi $ and $q(t)=q_{0}\cos t+p_{0}\sin t.\,\ $Due to the rotational
invariance of the harmonic oscillator Hamiltonian $H=(p^{2}+q^{2})/2$ we may
assume
\begin{equation}
q(t)=q_{0}\cos t;\quad p(t)=-q_{0}\sin t;\quad
p_{0}^{2}=2-q_{0}^{2}=0\Rightarrow q_{0}=\pm \sqrt{2}
\end{equation}
so that the classical tomogram $\mathcal{W}(X,1,0)$ is
\begin{equation}
\frac{1}{2\pi }\frac{2}{\left| q_{0}\sin (\arccos \frac{X}{q_{0}})\right| }%
\chi _{\lbrack -1,1]}(\frac{X}{q_{0}})=\frac{1}{\sqrt{2}\pi }\frac{1}{\sqrt{%
\left| 1-\left( \frac{X}{\sqrt{2}}\right) ^{2}\right| }}\chi _{\lbrack
-1,1]}(\frac{X}{\sqrt{2}})  \label{Woscclass}
\end{equation}
which coincides with the quantum Ehrenfest limit of Eq.(\ref{Wosc_infin}).
Furthermore,
\begin{equation}
dq(t)=-q_{0}\sin t\ dt\Rightarrow \left| \frac{dt}{dX}\right| =\frac{1}{%
\left| q_{0}\sin t\right| }=\frac{1}{\sqrt{2}}\frac{1}{\sqrt{\left| 1-\left(
\frac{X}{\sqrt{2}}\right) ^{2}\right| }}
\end{equation}
and $dt/dX$ is the inverse of the classical velocity $V$, so finally Eq.(\ref
{Woscclass}) can be written as:
\begin{equation}
\mathcal{W}(X,1,0)=\frac{2}{T}\frac{1}{\left| V\left( \frac{X}{\sqrt{2}}%
\right) \right| }\chi _{\lbrack -1,1]}(\frac{X}{\sqrt{2}}).
\end{equation}
As expected, the probability distribution of the position depends on the
inverse of the classical velocity $V$ in that point and diverges at the
turning points $X/\sqrt{2}=\pm 1$.

\section{Conclusions}

To conclude we found that the classical limit of quantum state tomograms is
suitable for a comparison with classical state tomograms. From the Planck
limit $\hbar \rightarrow 0,$ the tomograms of stationary states of the
harmonic oscillator, as well as of its coherent states, yield the localized
state $\delta (X),$ agreeing with the classical tomogram of a state
localized in the phase space. The Ehrenfest limit of quantum tomograms,
where the energy is fixed while $\hbar \rightarrow 0$, gives the expected
expressions of classical tomograms of classical states. When the energy
vanishes, the result is a rest state $\delta (X-\mu q_{0})$ with the
particle sitting at the minimum of the potential.

The same results were shown to apply in the case of both the Planck and the
Ehrenfest limits of the quantum tomogram for a particle in a box.

We have also found that, for the Schroedinger cat states, the interference
term of a superposition of two orthogonal states vanishes in the Planck
limit as $\sqrt{\hbar }$. In the Ehrenfest limit, even and odd coherent
states yield the expected mixture of two classical delta distribution.

For composite systems, we believe that tomographic description of quantum
and classical states can be a suitable tool to study the classical limit of
entangled states both in the Planck and in Ehrenfest limit. This aspect will
be considered in a forthcoming paper.

Finally we observe that the diverging inverse velocity factor in the
Ehrenfest limit of the harmonic oscillator results from the presence of
turning points of the harmonic oscillator motion along the $X-$axis. These
divergences should disappear when describing the motion with a complex
variable instead of a real one. To do that, the holomorphic (i.e.
Bargmann-Fock) representation is suitable. However, this requires a
preliminary discussion for dealing with tomograms associated with
non-Hermitian operators, which will appear elsewhere.

\section{Appendix}

Many results of the present paper follow from a Theorem, valid under quite
general assumptions, that here we recall:

\noindent \textbf{Dirac delta Theorem }Let $f$ be a summable function on the
real line such that
\begin{equation}
\int f(x)dx=N.
\end{equation}

Then $nf(n(x-x^{\prime }))\rightarrow N\delta (x-x^{\prime })$ when $%
n\rightarrow \infty .$

\noindent \textbf{Proof} Let $\varphi $ be any test function. Then by
Lebesgue Theorem
\begin{equation*}
\underset{n\rightarrow \infty }{\lim }n\int f(n(x-x^{\prime }))\varphi (x)dx=%
\underset{n\rightarrow \infty }{\lim }\int f(s)\varphi (\frac{s}{n}%
+x^{\prime })ds
\end{equation*}
\begin{equation}
=\int \underset{n\rightarrow \infty }{\lim }f(s)\varphi (\frac{s}{n}%
+x^{\prime })ds=\varphi (x^{\prime })\int f(s)ds=N\varphi (x^{\prime }).
\end{equation}

\noindent \textbf{Remark} We can evaluate the Planck limit of the product of
tomogram amplitudes $A_{\psi }A_{\phi }^{\ast },\,$when the state vectors $%
\left\langle y|\psi \right\rangle \ $and $\left\langle y|\phi \right\rangle $
have scaling exponent $\gamma =-1/2.$ In general, for any $\gamma ,\,$it is
\begin{eqnarray}
\int \frac{A_{\psi }A_{\phi }^{\ast }}{2\pi \hbar \left| \nu \right| }dX
&=&\int \psi (y)\phi ^{\ast }(u)e^{i\frac{\mu }{2\hbar \nu }%
(y^{2}-u^{2})}e^{-i\frac{X}{\hbar \nu }(y-u)}\frac{dydudX}{2\pi \hbar \left|
\nu \right| }  \notag \\
&=&\int \psi (y)\phi ^{\ast }(u)e^{i\frac{\mu }{2\hbar \nu }%
(y^{2}-u^{2})}\delta (y-u)dydu=\left\langle \phi |\psi \right\rangle
\end{eqnarray}
In the case$\ \gamma =-1/2,$ scaling the integration variables yields
\begin{equation}
\frac{A_{\psi }A_{\phi }^{\ast }}{2\pi \hbar \left| \nu \right| }=\int \Psi
(y)\Phi ^{\ast }(u)e^{i\frac{\mu }{2\nu }(y^{2}-u^{2})}e^{-i\frac{X}{\sqrt{%
\hbar }\nu }(y-u)}\frac{dydu}{2\pi \sqrt{\hbar }\left| \nu \right| }.
\end{equation}
So, as a consequence of the previous Theorem (with$\,n=1/\sqrt{\hbar }),$ it
results
\begin{equation}
\underset{\hbar \rightarrow 0}{\lim }\frac{A_{\psi }A_{\phi }^{\ast }}{2\pi
\hbar \left| \nu \right| }=\left\langle \phi |\psi \right\rangle \delta (X).
\end{equation}
For the tomogram amplitudes $A_{n}(X,\mu ,\nu )$ of the eigenstates $\psi
_{n}$ of a quantum harmonic oscillator, it results
\begin{equation}
\underset{\hbar \rightarrow 0}{\lim }\frac{A_{n}A_{m}^{\ast }}{2\pi \hbar
\left| \nu \right| }=\delta _{n,m}\delta (X).  \label{A19}
\end{equation}
where $\delta _{n,m}$ is the Kronecker delta.

When the scaling exponent of $\left\langle y|\psi \right\rangle ,$ $%
\left\langle y|\phi \right\rangle $\ is $\gamma =0,$ the same result can be
obtained trough their Fourier transforms which scale with $\gamma =-1/2.$

\end{document}